# STATISTICAL MECHANICS:

## A SELECTIVE REVIEW OF TWO CENTRAL ISSUES[*]

**Joel L. Lebowitz**

Departments of Mathematics and Physics

Rutgers University, Piscataway, New Jersey

E-mail: lebowitz@math.rutgers.edu

**Abstract**

I give a highly selective overview of the way statistical mechanics explains the microscopic origins of the time asymmetric evolution of macroscopic systems towards equilibrium and of first order phase transitions in equilibrium. These phenomena are emergent collective properties not discernible in the behavior of individual atoms. They are given precise and elegant mathematical formulations when the ratio between macroscopic and microscopic scales becomes very large.

## 1. Introduction

Nature has a hierarchical structure, with time, length and energy scales ranging from the submicroscopic to the supergalactic. Surprisingly it is possible and in many cases essential to discuss these levels independently—quarks are irrelevant for understanding protein folding and atoms are a distraction when studying ocean currents. Nevertheless, it is a central lesson of science, very successful in the past three hundred years, that there are no new fundamental laws, only new phenomena, as one goes up the hierarchy. Thus, arrows of explanations between different levels always point from smaller to larger scales,

---

[*] To appear in a special 1999 issue of Reviews of Modern Physics, celebrating the centennial of the American Physical Society

although the origin of higher level phenomena in the more fundamental lower level laws is often very far from transparent. (In addition some of the dualities recently discovered in string theory suggest possible arrows from the highest to the lowest level, closing the loop.)

Statistical Mechanics (SM) provides a framework for describing how well-defined higher level patterns or behavior may result from the nondirected activity of a multitude of interacting lower level individual entities. The subject was developed for, and has had its greatest success so far in, relating mesoscopic and macroscopic thermal phenomena to the microscopic world of atoms and molecules. Fortunately, many important properties of objects containing very many atoms—such as the boiling and freezing of water—can be obtained from simplified models of the structure of atoms and the laws governing their interactions. SM therefore often takes as its lowest level starting point—and so will I in this article— Feynman's description of atoms [1] as "little particles that move around in perpetual motion, attracting each other when they are a little distance apart, but repelling upon being squeezed into one another." Why this crude classical picture (a refined version of that held by some ancient Greek philosophers) gives predictions which are not only qualitatively correct but in many cases also highly accurate, is certainly far from clear to me—but that is another story or article.

Statistical mechanics explains how macroscopic phenomena originate in the cooperative behavior of these "little particles". Some of the phenomena are simple additive effects of the actions of individual atoms, e.g. the pressure exerted by a gas on the walls of its container, while others are paradigms of emergent behavior, having no direct counterpart in the properties or dynamics of individual atoms. Particularly fascinating and important examples of such emergent phenomena are the irreversible approach to equilibrium and phase transitions in equilibrium. Both of these would (or should) be astonishing if they were not so familiar. Their microscopic derivation and analysis forms the core of SM. I will discuss the first of these in section 2 and the second in section 3.

For a more general survey of SM in the past hundred years, the reader is referred to the other articles in this section as well as to my article in the special volume celebrating the first hundred years of the Physical Review [2] where there are also reprints of some of the original papers as well as references to others. For some very recent reviews of specific topics see [3].

## 2. Microscopic Origins of Irreversible Macroscopic Behavior

There are many conceptual and technical problems encountered in going from a time symmetric description of the dynamics of atoms to a time asymmetric description of the evolution of macroscopic systems. This involves a change from Hamiltonian (or Schrödinger) equations to hydrodynamical ones, e.g. the diffusion equation. The problem of reconciling the latter with the former became a central issue in physics during the last part of the nineteenth century. It was also in my opinion essentially resolved at that time, at least in the framework of nonrelativistic classical mechanics. To quote from Thomson's (later Lord Kelvin) 1874 article [4], "The essence of Joule's discovery is the subjection of physical phenomena to dynamical law. If, then, the motion of every particle of matter in the universe were precisely reversed at any instant, the course of nature would be simply reversed for ever after. The bursting bubble of foam at the foot of a waterfall would reunite and descend into the water .... Physical processes, on the other hand, are irreversible: for example, the friction of solids, conduction of heat, and diffusion. Nevertheless, the principle of dissipation of [organized] energy is compatible with a molecular theory in which each particle is subject to the laws of abstract dynamics." Unfortunately there is still much confusion about this issue among some scientists which is the reason for my discussing it here.[a]

Formally the problem considered by Thomson is as follows: The complete microscopic (or micro)state of an isolated classical system of $N$ particles is represented by a point $X$ in its phase space $\Gamma$, $X = (\mathbf{r}_1, \mathbf{p}_1, \mathbf{r}_2, \mathbf{p}_2, ..., \mathbf{r}_N, \mathbf{p}_N)$, $\mathbf{r}_i$ and $\mathbf{p}_i$ being the position and momentum of the $i$th particle. The evolution is governed by Hamiltonian dynamics, which connects a microstate at some time $t_0$, $X(t_0)$, to the microstate $X(t)$ at all other times $t$, $-\infty < t < \infty$. Let $X(t_0)$ and $X(t_0 + \tau)$, with $\tau$ positive, be two such microstates. Reversing (physically or mathematically) all velocities at time $t_0 + \tau$, we obtain a new microstate. If we now follow the evolution for another interval $\tau$ we find that the new microstate at time $t_0 + 2\tau$ is just $RX(t_0)$, the microstate $X(t_0)$ with all velocities reserved;

---

[a] This issue was the subject of a "round table" at the 20th IUPAP International Conference on Statistical Physics held in Paris, July 20–25, 1998. The panel consisted of M. Klein, who gave a historical overview, myself, who presented the Boltzmannian point of view described in the text which follows, I. Prigogine, who disagreed strongly with this point of view, claiming that the explanation lies in some (to me abstruse) new mathematical formalism developed by his group, and D. Ruelle, who presented some recent developments in the dynamical systems approach to far from equilibrium stationary states. The proceedings of that conference, which contain the presentations of the panel as well as some of the latest developments in SM, will appear in *Physica A*. See also [4f] and [4g].

$RX = (\mathbf{r}_1, -\mathbf{p}_1, \mathbf{r}_2, -\mathbf{p}_2, ..., \mathbf{r}_N, -\mathbf{p}_N)$. Hence, if there is an evolution (i.e. a trajectory $X(t)$), of a system in which some property of the system described by some function $f(X) = f(RX)$, which increases as $t$ increases, e.g. particle densities get more uniform by diffusion, there is also one in which the density profile evolves in the opposite direction, since the density is the same for $X$ and $RX$. So why is one direction, identified with "entropy" increase by the second "law", common and the other never seen?

The explanation of this apparent paradox, due to Thomson, Maxwell and Boltzmann, which I will now describe, shows that not only is there no conflict between reversible microscopic laws and irreversible macroscopic behavior, but, as clearly pointed out by Boltzmann in his later writings [b], there are extremely strong, albeit subtle, reasons to expect the latter from the former. These involve several interrelated ingredients which together provide the sharp distinction between microscopic and macroscopic variables required for the emergence of definite time asymmetric behavior in the evolution of the latter despite the total absence of such asymmetry in the dynamics of individual atoms. They are: a) the great disparity between microscopic and macroscopic scales, b) the fact that events are, as put by Boltzmann, determined not only by differential equations, but also by initial conditions, and c) the use of probabilistic reasoning: it is not every microscopic state of a macroscopic system that will evolve in accordance with the second law, but only the "majority" of cases—a majority which however becomes so overwhelming when the number of atoms in the system becomes very large that irreversible behavior becomes a

---

[b] Boltzmann's early writings on the subject are sometimes unclear, wrong, and even contradictory. His later writings, however, are superbly clear and right on the money (even if a bit verbose for Maxwell's taste). I strongly recommend the references cited at the end.

near certainty. (The characterization of the set whose "majority" we are describing will be discussed later.)

To see how the explanation works let us denote by $M$ the macrostate of a macroscopic system. For a system containing $N$ atoms in a box $V$, the microstate $X$ is a point in the $6N$ dimensional phase space $\Gamma$ while $M$ is a much cruder description, e.g. the specification, to within a given accuracy, of the energy of the system and of the number of particles in each half of the box. (A more refined (hydrodynamical) description would divide $V$ into $K$ cells, where $K$ is large, but still $K << N$, and specify the number of particles and energy in each cell, again with some tolerance.) Thus, while $M$ is determined by $X$ there are many $X$ which correspond to the same $M$. We will call $\Gamma_M$ the region in $\Gamma$ consisting of all microstates $X$ corresponding to a given macrostate $M$ and take as a measure of the "number" of microstates corresponding to a subset $A$ of $\Gamma_M$ to be equal to the $6N$ dimensional Liouville volume of $A$ normalized by the volume of $\Gamma_M$, denoted by $|\Gamma_M|$: $|\Gamma_M| = \int_{\Gamma_M} \Pi_{i=1}^N d\mathbf{r}_i d\mathbf{p}_i$. (This corresponds to the classical limit of "counting" states in quantum mechanics.)

Consider now a situation in which there is initially a wall confining a dilute gas of $N$ atoms to the left half of the box $V$. When the wall is removed at time $t_a$, the phase space volume available to the system is fantastically enlarged, roughly by a factor of $2^N$. (If the system contains 1 mole of gas in a container then the volume ratio of the unconstrained region to the constrained one is of order $10^{10^{20}}$). This region will contain new macrostates with phase space volumes very large compared to the initial phase space volume available to the system. We can then expect (in the absence of any obstruction, such as a hidden conservation law) that as the phase point $X$ evolves under the unconstrained dynamics it will with very high "probability" enter the newly available regions of phase space and thus find itself in a succession of new macrostates $M$ for which $|\Gamma_M|$ is increasing. This will continue until the system reaches its unconstrained macroscopic equilibrium state, $M_{eq}$, that is, until $X(t)$ reaches $\Gamma_{M_{eq}}$, corresponding to approximately half the particles in each half of the box, say within an interval $(\frac{1}{2} - \epsilon, \frac{1}{2} + \epsilon)$, $\epsilon << 1$, since in fact $|\Gamma_{M_{eq}}|/|\Sigma_E| \simeq 1$, where $|\Sigma_E|$ is the total phase space volume available under the energy constraint. After that time we can expect only small fluctuations about the value $\frac{1}{2}$, well within the precision $\epsilon$, typical fluctuations being of the order of the square root of the number of particles involved.

To extend the above observation to more general situations Boltzmann associated with each microscopic state $X$ of a macroscopic system, be it gas, fluid or solid, a number $S_B$, given, up to multiplicative and additive constants (in particular we set Boltzmann's constant, $k_B$, equal to unity), by

$$S_B(X) = \log |\Gamma_{M(X)}|. \qquad (2.1)$$

A crucial observation made by Boltzmann was that when $X \in M_{eq}$ then $S_B(X)$ agrees (up to terms negligible in the size of the system) with the thermodynamic entropy of Clausius and thus provides a microscopic definition of this macroscopically defined, operationally measurable (a la Carnot), extensive property of macroscopic systems in *equilibrium*. Having made this connection Boltzmann found it natural also to use (2.1) to define the entropy for a macroscopic system not in equilibrium and thus to explain (in agreement with the ideas of Maxwell and Thomson) the observation, embodied in the second law of thermody-namics, that when a constraint is lifted, an isolated macroscopic system will evolve toward a state with greater entropy[c], i.e. that $S_B$ will *typically* increase in a way which *explains* and describes qualitatively the evolution towards equilibrium of macroscopic systems.

Typical, as used here, means that the set of microstates corresponding to a given macrostate $M$ for which the evolution leads to a macroscopic decrease in the Boltzmann entropy during some fixed time period $\tau$, occupies a subset of $\Gamma_M$ whose Liouville volume is a fraction of $|\Gamma_M|$ which goes very rapidly (exponentially) to zero as the number of atoms in the system increases.

It is this very large number of degrees of freedom involved in the specification of macroscopic properties which distinguishes macroscopic irreversibility from the weak approach to equilibrium of ensembles for systems with good ergodic properties [4f]. While the former is manifested in a typical evolution of a single macroscopic system, the latter, which is also present in *chaotic* systems with but a few degrees of freedom, e.g. two hard spheres in a box, does not correspond to any appearance of time asymmetry in the evolution of an individual system. On the other hand, because of the exponential increase of the phase space volume, even a system with only a few hundred particles (commonly used in molecular dynamics computer simulations) will, when started in a nonequilibrium

---

[c] When $M$ specifies a state of local equilibrium, $S_B(X)$ agrees up to negligible terms, with the "hydrodynamic entropy". For systems far from equilibrium the appropriate definition of $M$ and thus of $S_B$ is more problematical. For a dilute gas in which $M$ is specified by the density $f(\mathbf{r},\mathbf{v})$ of atoms in the six dimensional position and velocity space $S_B(X) = -\int f(\mathbf{r},\mathbf{v}) \log f(\mathbf{r},\mathbf{v}) d\mathbf{r} d\mathbf{v}$. This identification is, however, invalid when the potential energy is not negligible; c.f. E. T. Jaynes, *Phys. Rev. A*, **4**, 747 (1971).

Following Oliver Penrose [5], we shall call $S_B(X)$ the Boltzmann entropy of the macrostate $M = M(X)$.

"macrostate" $M$, with 'random' $X \in \Gamma_M$, appear to behave like a macroscopic system.[d] This will be so even when integer arithmetic is used in the simulations so that the system behaves as a truly isolated one; when its velocities are reversed the system retraces its steps until it comes back to the initial state (with reversed velocities), after which it again proceeds (up to very long Poincare recurrence times) in the typical way [6].

Maxwell makes clear the importance of the scale separation when he writes [7]: "the second law is drawn from our experience of bodies consisting of an immense number of molecules. ... it is continually being violated, ..., in any sufficiently small group of molecules ... As the number ... is increased ...the probability of a measurable variation...may be regarded as practically an impossibility." We might take as a summary of the discussions in the late part of the last century the statement by Gibbs [8] quoted by Boltzmann (in a German translation) on the cover of his book *Lectures on Gas Theory II*: "In other words, the impossibility of an uncompensated decrease of entropy seems to be reduced to an improbability."

As already noted, typical here refers to a measure which assigns (at least approximately) equal weights to the different microstates consistent with the "initial" macrostate $M$. (This is also what was meant earlier by the 'random' choice of an initial $X \in \Gamma_M$ in the computer simulations.) In fact, any meaningful statement about probable or improbable behavior of a physical system has to refer to some agreed upon measure (probability distribution). It is, however, this use of probabilities (whose justification is beyond the reach of mathematical theorems) and particularly of the notion of typicality for explaining the origin of the apparently deterministic second law which was most difficult for many of Boltzmann's contemporaries, and even for some people today, to accept [4f, g]. This was clearly faced by Boltzmann when he wrote, in his second reply to Zermelo in 1897 [9] "The applicability of probability theory to a particular case cannot of course be proved rigorously. ... Despite this, every insurance company relies on probability theory. ... It is even more valid [here], on account of the huge number of molecules in a cubic millimetre... The assumption that these rare cases are not observed in nature is not strictly provable (nor is the entire mechanical picture itself) but in view of what has been said it is so natural and obvious, and so much in agreement with all experience with probabilities ... [that] ... It is completely incomprehensible to me how anyone can see a refutation of the applicability of probability theory in the fact that some other argument shows that exceptions must occur now and then over a period of eons of time; for probability theory itself teaches just the same thing."

---

[d] After all the likelihood of hitting, in the course of say one thousand tries, on something which has probability of order $2^{-N}$ is, for all practical purposes, the same, whether $N$ is a hundred or $10^{23}$.

It should be noted here that an important ingredient in the above analysis is the constancy in time, of the Liouville volume of sets in the phase space $\Gamma$ as they evolve under the Hamiltonian dynamics (Liouville's Theorem). Without this invariance the connection between phase space volume and probability would be impossible or at least very problematic. We also note that, in contrast to $S_B(X)$, the Gibbs entropy $S_G(\mu)$,

$$S_G(\mu) = -\int \mu \log \mu \, dX, \qquad (2.2)$$

is defined not for individual microstates but for statistical ensembles or probability distributions $\mu$. For equilibrium ensembles $S_G(\mu_{eq}) \sim \log |\Sigma_E| \sim S_B(X)$, for $X \in M_{eq}$, up to terms negligible in the size of the system. However, unlike $S_B$, $S_G$ does not change in

time even for time dependent ensembles describing (isolated) systems not in equilibrium. Hence the relevant entropy for understanding the time evolution of macroscopic systems is $S_B$ and not $S_G$.

**Initial Conditions**

Once we accept the statistical explanation of why macroscopic systems evolve in a manner that makes $S_B$ increase with time, there remains the nagging problem (of which Boltzmann was well aware) of what we mean by "with time": since the microscopic dynamical laws are symmetric, the two directions of the time variable are *a priori* equivalent and thus must remain so *a posteriori*.

Put another way: why can we use phase space arguments (or time asymmetric diffusion type equations) to predict the behavior of an *isolated* system in a nonequilibrium macrostate $M_b$ at some time $t_b$, e.g. a metal bar with a nonuniform temperature, in the future, i.e. for $t > t_b$, but not in the past, i.e. for $t < t_b$? After all, if the macrostate $M$ is invariant under velocity reversal of all the atoms, then the analysis would appear to apply equally to $t_b + \tau$ and $t_b - \tau$. A plausible answer to this question is to assume that the nonequilibrium state of the metal bar, $M_b$, had its origin in an even more nonuniform macrostate $M_a$, prepared by some experimentalist at some earlier time $t_a < t_b$ and that for states thus prepared we can apply our (approximately) equal a priori probability of microstates argument, i.e. we can assume its validity at time $t_a$. But what about events on the sun or in a supernova explosion where there are no experimentalists? And what, for that matter, is so special about the status of the experimentalist? Isn't he or she part of the physical universe?

Before trying to answer the last set of "big" questions let us consider whether the assignment of equal probabilities for $X \in \Gamma_{M_a}$ at $t_a$ permits the use of an equal probability distribution of $X \in \Gamma_{M_b}$ at time $t_b$ for predicting *future* macrostates: in a situation where the system is isolated for $t > t_a$. Note that the microstates in $\Gamma_{M_b}$, which have come from $\Gamma_{M_a}$ through the time evolution during the time interval from $t_a$ to $t_b$, make up only a very small fraction of the volume of $\Gamma_{M_b}$, call it $\Gamma_{ab}$. Thus we have to show that the overwhelming majority of points in $\Gamma_{ab}$ (with respect to Liouville measure on $\Gamma_{ab}$, which is the same as Liouville measure on $\Gamma_{M_a}$) have *future* macrostates like those typical of $\Gamma_b$—while still being very special and unrepresentative of $\Gamma_{M_b}$ as far as their *past* macrostates are concerned .[e] This property is explicitly proven by Lanford in his derivation of the Boltzmann equation (for short times) [10], and is part of the derivation of hydrodynamic equations [11]; see also [12].

To see intuitively the origin of this property we note that for systems with realistic interactions the domain $\Gamma_{ab}$ will be so convoluted as to *appear* uniformly smeared out in $\Gamma_{M_b}$. It is therefore reasonable that the future behavior of the system, as far as macrostates go, will be unaffected by their past history. It would of course be nice to prove this in all cases, e.g. justifying (for practical purposes) the factorization or "Stosszahlansatz" assumed by Boltzmann in deriving his dilute gas kinetic equation for all times $t > t_a$, not only for the short times proven by Lanford. Our mathematical abilities are, however, equal to this task only in very simple situations as we shall see below. This should, however, be enough to convince a 'reasonable' person.

---

[e] We are considering here the case where the macrostate $M(t)$, at time $t$, determines $M(t')$ for $t' > t$. There are of course situations where $M(t')$ depends also (weakly or even strongly) on the history of $M(t)$ in some time interval prior to $t'$, e.g. in materials with memory.

The large number of atoms present in a macroscopic system plus the chaotic nature of the dynamics also explains why it is so difficult, essentially impossible (except in some special cases such as experiments of the spin-echo type, and then only for a limited time), for a clever experimentalist to deliberately put such a system in a microstate which will lead it to evolve contrary to the second law. Such microstates certainly exist—just start with a nonuniform temperature, let it evolve for a while, then reverse all velocities. In fact, they are readily created in the computer simulations with no roundoff errors discussed earlier [6]. To quote again from Thomson's article [4]: "If we allowed this equalization to proceed for a certain time, and then reversed the motions of all the molecules, we would observe a disequalization. However, if the number of molecules is very large, as it is in a gas, any slight deviation from absolute precision in the reversal will greatly shorten the time during which disequalization occurs." In *addition*, the effect of unavoidable small outside influences, which are unimportant for the evolution of macrostates in which $|\Gamma_M|$ is increasing, will greatly destabilize evolution in the opposite direction when the trajectory has to *be aimed* at a very small region of the phase space [4f].

Let us return now to the big question posed earlier: what is special about $t_a$ compared to $t_b$ in a world with symmetric laws? Put differently, where ultimately do initial conditions such as those assumed at $t_a$ come from? In thinking about this we are led more or less inevitably to cosmological considerations and to postulate an initial "macrostate of the universe" having a very small Boltzmann entropy at some time $t_0$. To again quote Boltzmann [13]: "That in nature the transition from a probable to an improbable state does not take place as often as the converse, can be explained by assuming a very improbable [small $S_B$] initial [macro]state of the entire universe surrounding us. This is a reasonable assumption to make, since it enables us to explain the facts of experience, and one should not expect to be able to deduce it from anything more fundamental". We do not, however, have to assume a very special initial microstate $X$, and this is a very important aspect of our considerations. As Boltzmann further writes: "we do not have to assume a special type of initial condition in order to give a mechanical proof of the second law, if we are willing to accept a statistical viewpoint...if the initial state is chosen at random...entropy is almost certain to increase." All that is necessary to assume is a far from equilibrium initial macrostate and this is in accord with all cosmological and other independent evidence.

Feynman clearly agrees with this when he says [14], "it is necessary to add to the physical laws the hypothesis that in the past the universe was more ordered, in the technical sense, than it is today...to make an understanding of the irreversibility." More recently the same point was made very clearly by Roger Penrose in connection with the "big bang" cosmology [15]. Penrose, unlike Boltzmann, believes that we should search for a more fundamental theory which will also account for the initial conditions. Meanwhile he takes for the initial macrostate of the universe the smooth energy density state prevalent soon after the big bang. Whether this is the appropriate initial state or not, it captures an essential fact about our universe. Gravity, being purely attractive and long range, is unlike any of the other natural forces. When there is enough matter/energy around, it completely overcomes the tendency towards uniformization observed in ordinary objects at high energy densities or temperatures. Hence, in a universe dominated, like ours, by gravity, a uniform density corresponds to a state of very low entropy, or phase space volume, for a given total energy.

The local 'order' or low entropy we see around us (and elsewhere)—from complex molecules to trees to the brains of experimentalists preparing macrostates—is perfectly consistent with (and possibly even a consequence of) the initial macrostate of the universe. The value of $S_B$ of the present clumpy macrostate of the universe, consisting of planets, stars, galaxies, and black holes, is much much larger than what it was in the initial state and also quite far away from its equilibrium value. The 'natural' or 'equilibrium' state of the universe is, according to Penrose, one with all matter and energy collapsed into one big black hole which would have a phase space volume some $10^{10^{120}}$ times that of the initial macrostate. (So we may still have a long way to go.)

**Quantitative Considerations**

Let me now describe briefly the very interesting work, still in progress, in which one rigorously derives time asymmetric hydrodynamic equations from reversible microscopic laws [11]. While many qualitative features of irreversible macroscopic behavior depend very little on the positivity of Lyapunov exponents, ergodicity, or mixing properties of the microscopic dynamics, such properties are important for the quantitative description of the macroscopic evolution, i.e. for the derivation of time-asymmetric autonomous equations of hydrodynamic type. The existence and form of such equations depend on the instabilities of microscopic trajectories induced by chaotic dynamics. When the chaoticity can be proven to be strong enough (and of the right form) such equations can be derived rigorously from the reversible microscopic dynamics by taking limits in which the ratio of macroscopic to microscopic scales goes to infinity. Using the law of large numbers one shows that these equations describe the behavior of almost all individual systems in the ensemble, not just that of ensemble averages, i.e. that the dispersion goes to zero in the scaling limit. The equations also hold, to a high accuracy, when the macro/micro ratio is finite but very large.

A simple example in which this can be worked out in detail is the periodic Lorentz gas (or Sinai billiard). This consists of a *macroscopic number of non-interacting particles* moving among a periodic array of fixed convex scatterers, arranged in the plane in such a way that there is a maximum distance a particle can travel between collisions. The chaotic nature of the microscopic dynamics, which leads to an approximately isotropic local distribution of velocities, is directly responsible for the existence of a simple autonomous deterministic description, via a diffusion equation, for the macroscopic particle profiles of this system [11]. A second example is a system of hard spheres at very low densities for which the Boltzmann equation has been shown to describe the evolution of the density in the six dimensional position and velocity space (at least for short times) [10]. I use these examples, despite their highly idealized nature, because here all the mathematical i's have been dotted. They thus show *ipso facto*, in a way that should convince even (as Mark Kac put it) an "unreasonable" person, not only that there is no conflict between reversible microscopic and irreversible macroscopic behavior but also that, *for essentially all initial microscopic states consistent with a given nonequilibrium macroscopic state*, the latter follows from the former—in complete accord with Boltzmann's ideas. Yet the debate goes on.

## 3. Phase Transitions in Equilibrium Systems

Information about the equilibrium phases of a homogeneous macroscopic system is conveniently encoded in its phase diagram. Phase diagrams can be very complicated but their essence is already present in the familiar, simplified two dimensional diagram for a one component system like water or argon. This has axes marked by the temperature $T$ and pressure $p$, and gives the decomposition of this thermodynamic parameter space into different regions: the blank regions generally correspond to parameter values in which there is a unique pure phase, gas, liquid, or solid, while the lines between these regions represent values of the parameters at which two pure phases can exist. At the triple point, the system can exist in any of three pure phases.

In general, a macroscopic system with a given Hamiltonian is said to *undergo* or *be at* a first-order phase transition when the temperature and pressure or more generally the temperature and chemical potentials do not uniquely specify its homogeneous equilibrium state. The different properties of the pure phases coexisting at such a transition manifest themselves as discontinuities in certain observables, e.g., a discontinuity in the density as a function of temperature at the boiling point. On the other hand, when one moves between two points in the thermodynamic parameter space along a path which does not intersect any coexistence line the properties of the system change smoothly.

I will now sketch a mathematically precise formulation of what is meant by coexistence of phases, and give some rigorous results about phase diagrams. This is a beautiful part of the developments in SM during this century, it is also one which is essential to a full understanding of the singular behavior of macroscopic systems at phase transitions, e.g.

the discontinuity in the density mentioned earlier. These singularities can only be captured precisely through the infinite volume or *thermodynamic limit* (TL); a formal mathematical procedure in which the size of the system becomes infinite while the number of particles and energy per unit volume (or the chemical potential and temperature) stay fixed. While at first sight entirely unrealistic, such a limit represents an idealization of a macroscopic physical system whose spatial extension, although finite, is very large on the microscopic scale of interparticle distances or interactions. The advantage of this idealization is that boundary and finite size effects present in real systems, which are frequently irrelevant to the phenomena of interest, are eliminated in the TL. As Robert Griffiths once put it, every experimentalist implicitly takes such a limit when he or she reports the results of a measurement, like the magnetic susceptibility, without giving the size and shape of the sample.

My starting point here is the Gibbs formalism for calculating equilibrium properties of macroscopic systems as ensemble averages of *functions* of the microscopic state of the system. While the use of ensembles was anticipated by Boltzmann [4d,8b] and independently discovered by Einstein, it was Gibbs who, by his brilliant systematic treatment of statistical ensembles, i.e. probability measures on the phase space, developed SM into a useful elegant tool for relating, not only typical but also fluctuating behavior in equilibrium systems, to microscopic Hamiltonians. In a really remarkable way the formalism has survived essentially intact the transition to quantum mechanics. Here, however, I restrict myself to classical mechanics.[f]

As in section 2, the microscopic state of a system of $N$ particles in a spatial domain $V$ is given by a point $X$ in the phase space, $X = (\mathbf{r}_1, \mathbf{p}_1, ..., \mathbf{r}_N, \mathbf{p}_N)$. We are generally interested in the values of suitable *sum functions* of $X$: those which can be written as a sum of terms involving only a fixed finite number of particles, e.g. $F_{(1)}(X) = \sum f_1(\mathbf{r}_i, \mathbf{p}_i)$, $F_{(2)}(X) = \sum_{i,j} f_2(\mathbf{r}_i, \mathbf{p}_i, \mathbf{r}_j, \mathbf{p}_j)$, (with $f_2(\mathbf{r}_i, \mathbf{p}_i, \mathbf{r}_j, \mathbf{p}_j) \to 0$ when $|\mathbf{r}_i - \mathbf{r}_j| \to \infty$), etc. (Familiar examples are the kinetic and potential energies of the system.) Typical macroscopic properties then correspond to sum functions which, when divided by the volume $|V|$, are essentially constant on the energy surface $\Sigma_E$ of a macroscopic system. Consequently, if we take the TL, defined by letting $N \to \infty$, $E \to \infty$, and $|V| \to \infty$ in such a way that $N/|V| \to \rho$ and $E/|V| \to e$, then these properties assume deterministic values, i.e. their variances go to zero. They also become (within limits) independent of the shape of $V$ and the nature of the boundaries of $V$. (As a less familiar concrete example, let $f_1(\mathbf{r}_i, \mathbf{p}_i) = (\frac{1}{2m}\mathbf{p}_i^2)^2$, the square of the kinetic energy of the $i$th particle. Then, in the TL, $|V|^{-1} F_1(X) \to \frac{9}{4}\rho T^2(e, \rho)$ for *typical* $X$, with $T$ the temperature of the system given by $[\frac{\partial}{\partial e} s(e, \rho)]^{-1}$, with $s(e, \rho)$ the TL of $|V|^{-1} \log |\Sigma_E|$).

It is this property of sum functions which makes meaningful the use of ensembles to describe the behavior of individual macroscopic systems as in section 2. In particular it assures the "equivalence" of ensembles: microcanonical, canonical, grand canonical, pressure, etc. for computing equilibrium properties. The use of the TL actually extends this equivalence, in that part of the phase diagram where the system has a unique phase, to the probability distribution of fluctuating quantities, e.g. the correlation functions. These are translation invariant and independent of boundary conditions in the TL. (See later for what happens on coexistence lines.)

---

[f] It is clearly impossible to cite here all or even a significant fraction of all the good reviews and textbooks on the subject. The reader would do well however to browse among the original works [4, 5] and in particular read Gibbs' beautiful book [16]. A partial list of books and reviews with a mathematical treatment of Gibbs measures and phase transitions which contain the results presented without references can be found in ref. [17].

To actually obtain the phase diagram of a system with a given Hamiltonian is a formidable mathematical task. It has still not been solved even for such simple continuum systems as particles interacting via a Lennard-Jones pair potential. I will therefore postpone further discussion of continuum systems until later and switch now to lattice systems for which such results are available. These come from a variety of techniques some of which, I shall not be able to discuss here at all.

**Lattice Systems**

Lattice systems can be considered approximations to the continuum particle systems (the cell theory of fluids) or as representations of spins in magnetic systems [18]. They also arise as models of a variety of non-thermal physical phenomena [19]. I shall consider for simplicity the simple cubic lattice, $\mathbb{Z}^d$, in $d$ dimensions. At each site $\mathbf{x} \in \mathbb{Z}^d$ there is a spin variable $S(\mathbf{x})$ which can take $k$ discrete values, $S(\mathbf{x}) = \xi_1, ..., \xi_k$. The configuration of the system in a region $V \subset \mathbb{Z}^d$ containing $|V|$ sites, is denoted by $\mathbf{S}_V$, it is one of the $k^{|V|}$ points in the set $\Omega = \{\xi_1.,,,,\xi_k\}^V$. There is an interaction energy $U$ which is a sum of *internal* interactions assumed to be translation invariant and *boundary terms*.

To be specific, consider the Ising model, $S(\mathbf{x}) = \pm 1$, with uniform magnetic field $h$

and pair interactions $u(\mathbf{r})$. The energy of a configuration $\mathbf{S}_V$ is given by,

$$U(\mathbf{S}_V|\bar{S}_{V^c}) = -h\sum_{\mathbf{x}\in V} S(\mathbf{x}) - \frac{1}{2}\sum_{\mathbf{x},\mathbf{y}\in V}\sum u(\mathbf{x}-\mathbf{y})S(\mathbf{x})S(\mathbf{y}) - \sum_{\mathbf{x}\in V}\{\sum_{\mathbf{y}\in V^c} u(\mathbf{x}-\mathbf{y})\bar{S}(\mathbf{y})\}S(\mathbf{x}). \tag{3.1}$$

In (3.1) $\bar{S}(\mathbf{y})$ denotes the *preassigned* value of the spin variables at sites $\mathbf{y}$ in $V^c$, the complement (or outside) of $V$, which act as boundary conditions (bc). They contribute, through the last sum in (3.1), an energy term which is proportional to the surface area of $V$ whenever the interactions have finite range or decay fast enough to be summable, e.g. $u(\mathbf{r})$ decays faster than $|\mathbf{r}|^{-(d+\epsilon)}$, $\epsilon > 0$. We can also consider periodic or free bc: the latter corresponds to dropping the last term in (3.1). We will indicate all possible boundary conditions by the letter $b$; sometimes setting $b = p$ or $b = f$ for periodic or free bc.

When the system is in equilibrium at temperature $T$, the probability of finding the configuration $\mathbf{S}_V$ is given by the Gibbs formula [17]

$$\mu_V(\mathbf{S}_V|b) = \frac{1}{Z(\mathbf{J};b,V)}\exp[-\beta U(\mathbf{S}_V|b)] \tag{3.2}$$

where $\beta^{-1} = T$, and $Z$ is the partition function,

$$Z(\mathbf{J};b,V) = \sum_{\mathbf{S}_V}\exp[-\beta U(\mathbf{S}_V|b)]. \tag{3.3}$$

The sum in (3.3) is over all possible microscopic configurations of the system in $V$ and we have used $\mathbf{J}$ to refer to all the parameters entering $Z$ through the interactions (including $\beta$) while $b$ represents the bc specified by $\bar{S}_{V^c}$, $p$, or $f$. The Gibbs free energy density of the finite system is given by

$$\Psi(\mathbf{J};b,V) \equiv |V|^{-1}\log Z(\mathbf{J};b,V) \tag{3.4}$$

To get an *intrinsic* free energy, which determines the bulk properties of a macroscopic system, one needs to let the size of $V$ become infinite while keeping $\mathbf{J}$ fixed in such a way that the ratio of surface area to volume goes to zero, i.e. to take the TL, $V \nearrow \mathbb{Z}^d$, in (3.4).

It is one of the most important rigorous results of SM, to whose proof many have contributed (see references [17a,b,c]) that when the interactions decay in a summable way, the limit $V \nearrow \mathbb{Z}^d$ of (3.5) in fact exists and is independent of the boundary condition $b$

$$\Psi(\mathbf{J}; b, V) \to \Psi(\mathbf{J}). \tag{3.5}$$

We shall call $\Psi(\mathbf{J})$ the thermodynamic free energy density. It has all the convexity properties of the free energy *postulated* by macroscopic thermodynamics as a stability requirement on the equilibrium state. (For Coulomb interactions see below and [3b]).

We now note that as long as $V$ is finite, $Z(\mathbf{J}; b, V)$ is a finite sum of positive terms and so $\Psi(\mathbf{J}; b, V)$ is a smooth function of the parameters $\mathbf{J}$ (including $\beta$ and $h$) entering the interaction. This is also true for the probabilities of the spin configuration in a set $A \subset V$, $\mu_V(\mathbf{S}_A|b)$ obtained from the Gibbs measure (3.2) or equivalently the correlation functions. In other words, once $b$ is specified, all equilibrium properties of the finite system vary smoothly with the parameters $\mathbf{J}$. The only way to get non-smooth behavior of the free energy or nonuniqueness of the measure is to take the TL. In that limit the $b$-independent $\Psi(\mathbf{J})$ can indeed have singularities. Similarly, the measure defined by a specification of the probabilities in *all* fixed regions $A \subset \mathbb{Z}^d$, $\hat{\mu}(\mathbf{S}_A|\hat{b})$, can depend on the way in which the TL was taken and in particular on the boundary conditions at "infinity", here denoted symbolically by $\hat{b}$ [17].

To see this explicitly, let us specialize even further and consider isotropic nearest neighbor (nn) interactions

$$u(\mathbf{r}) = \begin{cases} J, & \text{for } |\mathbf{r}| = 1 \\ 0, & \text{otherwise} \end{cases} \quad (3.6)$$

with $J$ constant. For this model the effect of the spins outside $V$, $\bar{S}_{V^c}$, is just to produce an additional magnetic fields $h_b(\mathbf{x})$, for $\mathbf{x}$ on the inner boundary of $V$. The finite volume free energy $\Psi(J_1, J_2; b, V)$, where $\beta h = J_1$ and $\beta J = J_2$, is then clearly real analytic for all $J_1, J_2 \in (-\infty, \infty)$. The phase diagram of this system after taking the TL is given in Fig. 1 where we have used axes labeled by $h/|J|$ and $J_2^{-1}$. Note that $J_2 > 0$ ($J_2 < 0$) corresponds to ferromagnetic (antiferromagnetic) interactions.

For the ferromagnetic Ising model, corresponding to the upper half of this figure, *almost* everything is known rigorously. In the region where the magnetic field $h$ is not zero, both $\Psi(J_1, J_2)$ and the infinite volume Gibbs measure, i.e. the $\hat{\mu}(\mathbf{S}_A|\hat{b})$, are independent of the bc and are real analytic in $J_1$ and $J_2$. The analyticity results follow from the remarkable Lee-Yang theorem [18] which states that for $J_2 \geq 0$ fixed, and $b = p$ or $f$, the only singularities of $\Psi(J_1, J_2 : b, V)$ (corresponding to zeros of the partition function) in the complex $J_1$ plane occur on the line $\text{Re} J_1 = 0$. Uniqueness of $\hat{\mu}$ follows [20] from an argument combining the Lee-Yang theorem with the equally remarkable Fortuin, Kasteleyn, Ginibre (FKG) inequalities [21].

Furthermore, for small values of $|J_2|$, $\Psi$ is analytic in both $J_1$ and $J_2$ and the measure $\hat{\mu}$ is unique. This fact, which holds for general interactions at high temperatures, follows either from the existence of a convergent high temperature expansion for $\Psi$ and for the correlation functions in powers of $\beta$ or from the Dobrushin-Shlosman uniqueness criterion [23]. On the other hand for $J_1 = 0$ and $J_2$ large enough there is the ingenious argument due to Peierls [22], made fully rigorous by Dobrushin and by Griffiths [17b,c], which proves that in dimension $d \geq 2$, the probability that the spin $S(\mathbf{x})$ has value $+1$ is different for "$b = +$" and "$b = -$", corresponding to bc for which $\bar{S}(\mathbf{y}) = +1$, or $\bar{S}(y) = -1$, respectively, for all $\mathbf{y}$ outside $V$. The crucial point of the Peierls argument is that this difference persists *no matter how large $V$* is: the probability being greater (less) than $\frac{1}{2}$ for $+$ $(-)$ b.c. This implies that the average value of the magnetization is positive at low temperatures for $+$ bc, even when $h = 0$, independent of $V$. By symmetry the opposite is true for $-$ bc. Thus for $J_1 = 0$ and $J_2$ large, the limiting Gibbs measures $\hat{\mu}_+$ and $\hat{\mu}_-$ (obtained with $+$ or $-$ bc), which can be shown to exist, are different. *It is this nonuniqueness of the Gibbs measure $\hat{\mu}$, for specified $\mathbf{J}$, which corresponds to the coexistence of phases in macroscopic systems.*

The expected value of $S(\mathbf{x})$ in the "$+$ state", denoted by $m^*(\beta)$, is independent of $\mathbf{x}$ and is equal to the value of the average of the magnetization in all of $V$ obtained when one lets $h \to 0$ from the positive side after taking the TL. (Remember that $\hat{\mu}$ and hence the magnetization, $m(\beta, h)$, is independent of b.c. for $h \neq 0$). It can be further shown, using the second Griffiths inequality that $m^*(\beta)$ is monotone increasing in $\beta$ [24]. Hence there is, for a given $J_2 > 0$, a unique critical temperature, $T_c$, such that for $h = 0$ and $T < T_c$, $m^*(\beta) > 0$ while for $T > T_c$, $m^*(\beta) = 0$. $T_c$ depends on the dimension $d$, $T_c(d) > 0$ for $d \geq 2$, $T_c(1) = 0$.

There is a unique infinite volume Gibbs measure for $T \geq T_c$ and (essentially) only two, $\hat{\mu}_+$ ands $\hat{\mu}_-$, *extremal, translation invariant* (TI) Gibbs measures for $T < T_c$. The latter statement means that every infinite volume TI Gibbs measure $\hat{\mu}_b$ is a convex combination of $\hat{\mu}_+$ and $\hat{\mu}_-$, i.e.

$$\hat{\mu}(\mathbf{S}_A|\hat{b}) = \alpha\hat{\mu}_+(\mathbf{S}_A) + (1-\alpha)\hat{\mu}_-(\mathbf{S}_A), \tag{3.7}$$

for some $\alpha$, $0 \leq \alpha \leq 1$. For periodic or free bc $\alpha = \frac{1}{2}$ by symmetry, so that $\hat{\mu}_p = \hat{\mu}_f = \frac{1}{2}(\hat{\mu}_+ + \hat{\mu}_-)$. This means physically that when $V$ is large the system with "symmetric" bc will, with equal probability, be found in *either* the "+ state" *or* in the opposite "− state". Of course as long as the system is finite it will "fluctuate" between these two pure phases, but the "relaxation times" for such fluctuations grows (for any reasonable dynamics) exponentially in $|V|$, so the *either/or* description correctly captures the behavior of macroscopic systems. This phenomena is the paradigm of *spontaneous symmetry breaking* which occurs in many physical situations.

The fact that free bc lead to translation invariant measures is a consequence of the Griffiths inequalities [24]. There also exist non-translation invariant $\hat{\mu}$ for temperatures below the "roughening" temperature $T_R \leq T_c$ [17]. These are obtained as the TL of systems with 'Dobrushin bc' favoring an interface between the + and − phase. Dobrushin [25] proved that $T_R > 0$ in $d \geq 3$ while Aizenman showed that long wavelength fluctuations destroy these states in two dimensions at all $T > 0$, i.e. $T_R = 0$ in $d = 2$ [26]. Using inequalities van Beijeren showed that $T_R(d) \geq T_c(d-1)$ [27].

We also know that at $T_c$, $m^*(\beta_c) = 0$ in $d = 2$ and for $d \geq 4$; the former from Onsager's exact solution [28] and the latter from general results about mean field like behavior for $d > 4$ [29] (with logarithmic corrections for $d = 4$). Of course one expects continuity of $m^*(\beta)$ for this system also in $d = 3$, but this is not yet proven. As the temperature is lowered, the + and − states come to resemble the two ground states corresponding to all spins up or all spins down, and there is a convergent low temperature "cluster expansion", in which the low order terms correspond to excitations consisting of small isolated domains of down (up) spins in the $+$ $(-)$ phase.

The absence of any homogeneous pure phases other than $\hat{\mu}_+$ and $\hat{\mu}_-$ i.e. the validity of (3.7) for all $TI$ $\hat{\mu}$, is only proven subject to the condition that the average energy is a continuous function of the temperature [30]. This is known in $d = 2$ from Onsager's solution which also gives the exact value of $T_c$. In $d > 2$ the continuity of the energy is known to hold at low temperatures (where the cluster expansion is valid) and at *almost all* temperatures otherwise. There is, however, much numerical and analytic evidence that $\Psi(J_1 = 0, J_2)$ is real analytic in $J_2$ everywhere away from the critical temperature. The story is similar for the decay of correlations. This is known to be exponential for $h \neq 0$ at high temperatures and at low temperatures in the $+$ and $-$ phases. Similar behavior is expected at all $T \neq T_c$, but this is only proven for $d = 2$ (and in $d = 1$ where $T_c = 0$). Note that for mixed states, when $\alpha \neq 0$ or 1 in (3.7), there is no decay of correlations.

Essentially everything said above for the ferromagnetic Ising model with nn interactions holds also for more general ferromagnetic pair interactions, $u(\mathbf{r}) \geq 0$ in (3.1) with $u(\mathbf{r})$ of finite range or decaying faster than $r^{-(d+1+\epsilon)}$. (An exception is the decay of correlations, which is never faster than the decay of the interactions). It follows in fact from the Griffiths, Kelley, Sherman (GKS) inequalities [24] that adding ferromagnetic pair or multi-spin interactions to an already ferromagnetic Ising system (with $h \geq 0$) can only increase the magnetization. A particular consequence of this is that the critical temperature for the nearest neighbor Ising model cannot decrease with dimension: going from $d$ to $d+1$ can be viewed of as adding ferromagnetic couplings. This argument works also when we increase the "thickness" of a $d$-dimensional system, e.g. adding layers to a $d=2$ Ising model. To show that $T_c$ actually increases, not just stay fixed, is more difficult. In fact, going from $d=1$ to a strip of finite width (and infinite length) does not increase $T_c$ from zero, its value for $d=1$.

An interesting situation occurs in $d=1$ when the ferromagnetic pair interaction decays like $r^{-\gamma}$, $1<\gamma\leq 2$, so the TL of $\Psi$ still exists. The $d=1$ system then has a $T_c>0$ with the spontaneous magnetization *discontinuous* at $T_c$ (coexistence of phases) for the borderline case $J(r)\sim r^{-2}$. A result of this type was first found by Yuval and Anderson, then proven for the hierarchical model by Dyson and for the Ising model by Aizenman et al. [31].

For general lattice systems we still have the existence of the TL of the free energy, independent of bc as well as the general connection between pure phases and extremal TI or periodic Gibbs states. (Any periodic Gibbs state can be made TI by enlarging the "unit cell" of the lattice.) We know, however, much less about the phase diagram, except at very low temperatures. Here the Pirogov-Sinai theory and its extensions [32] show how the existence of different periodic ground states, corresponding to a "ground states" diagram in the space of interactions **J**, at $T=0$, gives rise to a similar phase-diagram of the pure phases at sufficiently low temperatures. The great advantage of this theory, compared to arguments of the Peierls type, is that there is no requirement of symmetry between the phases—the existence of which was of crucial importance for the ferromagnetic examples discussed earlier.

This can be seen already for the $nn$ Ising model with antiferromagnetic interaction, $J < 0$ in (3.6). For $h = 0$ this system can be mapped into the ferromagnetic one by changing $S(\mathbf{x})$ into $-S(\mathbf{x})$ on the odd sublattice—but what about $h \neq 0$? If we look at this system at $T = 0$ we find two periodic ground states for $|h| < 2d|J|$ corresponding to $S(\mathbf{x}) = -1$ on the even (odd) and $S(\mathbf{x}) = 1$ on the odd (even) sublattice. For $|h| > 2d|J|$ there is a unique ground state: all up for $h > 2d|J|$, all down for $h < -2d|J|$. At $|h| = 2d|J|$ there are an infinite number of ground states with positive entropy per site (in violation of the third law). The existence of two periodic phases for sufficiently low temperatures at $|h| < 2d|J|$, and of a unique TI phase for $|h| > 2d|J|$ then follows from P-S theory. Of course for $h = 0$ we know, from the isomorphism with the ferromagnetic system, that there are two periodic states for all $T < T_c$. This, however, doesn't strictly (i.e. rigorously) tell us anything about $h \neq 0$. I am not aware of any argument which proves that the boundary of the curve enclosing the coexistence region in the antiferromagnetic part of Fig. 1 has to touch the point corresponding to $h = 0$, $T = T_c$. There is also for this system, a generalization of the Peierls argument, due to Dobrushin [33], which exploits the symmetry of this system and is therefore simpler than P-S theory. This proves the existence of the two periodic states in a portion of the phase diagram (indicated by the solid curve in Fig. 1).

Unfortunately P-S theory doesn't say anything about the immediate neighborhood of the points $|h| = 2d|J|$ where the system has such a high degeneracy. It does follow however from the general Dobrushin-Shlosman uniqueness criteria [23], implemented by computer enumerations, that, for $d = 2$, the boundary of the coexistence region at $4|J|$ has to curve to the left, hence we expect that for $|h| = 4|J|$ there is a unique phase for all $T > 0$.

**Continuum Systems**

The existence of the TL of the free energy, independent of $b$ as in (3.5), also holds for continuum systems (classical or quantum) with Hamiltonians, having the form

$$H(X) = (2m)^{-1} \sum \mathbf{p}_i^2 + \sum_{i \neq j} u(\mathbf{r}_i - \mathbf{r}_j), \tag{3.8}$$

and satisfying certain conditions [17]. These conditions are readily shown to hold for systems with Lennard-Jones type potentials. For Coulomb systems, where $u$ contains explicitly terms of the form $e_{\alpha_i} e_{\alpha_j} |\mathbf{r}_i - \mathbf{r}_j|^{-1}$ in $d = 3$ (logarithmic ones in $d = 2$, etc.) it is required that the system be overall charge neutral. For classical systems it is further required that there be some cutoff preventing arbitrarily large negative "binding" energies between positive and negative charges, e.g. a hard core exclusion. For quantum systems it is sufficient if either the positive or negative charges obey Fermi statistics—as electrons indeed do, see [34], [3b] and references there.

Remarkably enough it is possible to prove (subject to some assumptions) that a system of protons and electrons will, in certain regimes of sufficiently low temperatures and densities, consist mostly of a gas of atoms or molecules in their ground states [3b]. This may be the *beginning* of a theory which would justify, from first principles, the use of effective potentials, e.g. Feynman's "little particles" [1], for obtaining properties, such as phase transitions, of macroscopic systems; see also Fisher's intriguing article in [35].

I return now to the theme of this section with a discussion of first order phase transitions in continuum systems, a subject of much curent interest to me. While the general theory concerning infinite volume Gibbs measures readily extends to such systems, the techniques used for proving existence of phase transitions in lattice systems are harder to generalize. The ground states of even the simplest model continuum systems are difficult to characterize; they are presumed to be periodic or quasi-periodic configurations which depend in some complicated way on the interparticle forces. This is however far from proven, and hence the analysis of the fluctuations that appear when we increase the temperature above zero is correspondingly harder, indeed very much harder, to study than in lattice systems. Moreover, key inequalities are no longer available. These problems have been overcome for some multicomponent systems with special features. In particular Ruelle [36] proved that the symmetric two component Widom-Rowlinson model [37] has a demixing phase transition in $d \geq 2$. There are also later proofs of phase transitions in $d \geq 2$ for generalizations of this model as well as for $d = 1$ continuum systems with interactions which decay very slowly [38].

The first proof of a liquid-vapor transition in a one-component continuum systems with finite range interactions and no symmetries was given only very recently [39]. The basic idea there is to study perturbations not of the ground state but of the mean field state which describes systems with infinite range interactions. These interactions are parametrized by their range $\gamma^{-1}$ and the perturbation is about $\gamma = 0$. The proof of mean field behavior, in the limit $\gamma \to 0$, was first given in [40] for $d = 1$. These results were later generalized [41] to $d$-dimensional systems with suitable short range repulsive interactions and general Kac potentials of the form

$$\phi_\gamma(q_i, q_j) = -\alpha \gamma^d J(\gamma |q_i - q_j|), \qquad (3.9)$$

with $\int_{\mathbb{R}^d} J(r) dr = 1$, $J(r) > 0$. In the TL, followed by the limit $\gamma \to 0$, the Helmholtz free energy $a$ takes a mean field form,

$$\lim_{\gamma \to 0} a(\rho, \gamma) = CE\{a_0(\rho) - \frac{1}{2}\alpha\rho^2\}. \tag{3.10}$$

Here $\rho$ is the particle density and $a_0$ is the free energy density of the reference system, i.e. the system with no Kac potential. $a_0$ is convex in $\rho$ (by general theorems) and $CE\{f(x)\}$ is the largest convex lower bound of $f$. For $\alpha$ large enough the term in the curly brackets in (3.10) has a double well shape and the $CE$ corresponds to the Gibbs double tangent construction. This is equivalent to Maxwell's equal area rule applied to a van der Waals' type equation of state where it gives the coexistence of liquid and vapor phases. In this limit, $\gamma \to 0$, the correlation functions in the pure phases are those of the reference system at the corresponding densities.

The assumption of strongly repulsive short range interactions in [40,41], in addition to the long range attractive Kac type interactions, was dictated not only by realism but also by the need to insure stabilization against collapse, which would be induced by a purely attractive pair potential. The approach in [39], however, which proves a liquid-vapor phase coexistence for $\gamma > 0$, needs a cluster expansion for the unperturbed reference system (i.e. without the Kac interaction) at values of the chemical potential or density for which it is not proven to hold in systems with strong short range interactions. Stability is therefore produced by a positive four body potential of the same range as the attractive two body one. The reference system is then the free, ideal gas for which the cluster expansion holds trivially. The proof of the existence of phase-transitions in fluids with Lennard-Jones type potentials is therefore still an open problem. Hopefully we will not have to wait another century for its resolution.

## Acknowledgments

I would like to thank S. Goldstein and E. Speer for very many, very helpful comments. Thanks are also due to the anonymous referee for many suggestions, and to D. Chelst for a careful reading of the manuscript. The research was supported in part by NSF Grant 95-23266, and AFOSR Grant 95-0159.

## Figure Caption

**Fig. 1** Schematic phase diagram of nearest neighbor Ising model on a simple cubic lattice in dimensions $d \geq 2$. The ground states of the antiferromagnetic system are degenerate for $|h| \leq 2|J|d$. For $d = 1$, $T_c = 0$. The region above the dashed line is the expected two phase region. The one above the solid line is the proven one.

## References

[1] R.P Feynman, R. B Leighton, M. Sands, *The Feynman Lectures on Physics*, Addison-Wesley, Reading, Mass. (1963), section 1–2.

[2] J. L. Lebowitz, Introduction to Chapter 6, *Phys. Rev. Centennial*, in *The Physical Review: The First Hundred Years*, H. Stroke (ed.), AIP Press, p. 369, New York, 1995.

[3a] M. E. Fisher, Rev. Mod. Phys. **70**, 653 (1998).
[3b] D. Brydges and Ph. A. Martin, Rev. Mod. Phys., to appear.

[4a] W. Thomson, Proc. of the Royal Soc. of Edinburgh, **8**, 325 (1874).
[4b] Thomson's article as well as many other original articles from the second half of the nineteenth century on this subject (all in English) are reprinted in S.G. Brush, *Kinetic Theory*, Pergamon, Oxford, (1966).
[4c] For an interesting biography of Boltzmann, which also contains many references, see E. Broda *Ludwig Boltzmann, Man—Physicist—Philosopher*, Ox Bow Press, Woodbridge, Conn (1983); translated from the German.

[4d] For a historical discussion of Boltzmann and his ideas see articles by M. Klein, E. Broda, L. Flamn in *The Boltzmann Equation, Theory and Application*, E.G.D. Cohen and W. Thirring, eds., Springer-Verlag, 1973.

[4e] For a general history of the subject see S.G. Brush, *The Kind of Motion We Call Heat*, Studies in Statistical Mechanics, vol. VI, E.W. Montroll and J.L. Lebowitz, eds. North-Holland, Amsterdam, (1976).

[4f] For some of my own articles on this subject see J. L. Lebowitz, *Physica A* **194**, 1–97 (1993); Boltzmann's Entropy and Time's Arrow, *Physics Today*, **46**, 32–38 (1993); Lebowitz's replies to readers' responses to the previous article, Physics Today, 47:113–116, 1994; Microscopic Reversibility and Macroscopic Behavior: Physical Explanations and Mathematical Derivations, in *25 Years of Non-Equilibrium Statistical Mechanics*, Proceedings, Sitges Conference, Barcelona, Spain, 1994, in Lecture Notes in Physics, J.J. Brey, J. Marro, J.M. Rubí and M. San Miguel eds., Springer, (1995).

[4g] For a clear defense of Boltzmann's views against some recent attacks see the article by J. Bricmont, Science of Chaos or Chaos in Science? in *The Flight from Science and Reason*, Ann. N.Y. Academy of Sciences **79**, p. 131 (1996). This article first appeared in the publication of the Belgian Physical Society, Physicalia Magazine, **17**, 159, (1995), where it is followed by an exchange between Prigogine and Bricmont.

[5] O. Penrose, *Foundations of Statistical Mechanics*, Pergamon, Elmsford, N.Y. (1970), ch. 5.

[6] D. Levesque and L. Verlet, J. Stat. Phys. **72**, 519 (1993); see also B.T. Nadiga, J.E. Broadwell and B. Sturtevant, *Rarefield Gas Dynamics: Theoretical and Computational Techniques*, edited by E.P. Muntz, D.P. Weaver and D.H. Campbell, Vol 118 of *Progress in Astronautics and Aeronautics*, AIAA, Washington, DC, ISBN 0–930403–55–X, 1989.

[7] J.C. Maxwell, *Theory of Heat*, p. 308: "Tait's Thermodynamics", Nature **17**, 257 (1878). Quoted in M. Klein, ref. 4d).

[8a] J.W. Gibbs, Connecticut Academy Transactions **3**, 229 (1875), reprinted in *The Scientific Papers*, **1**, 167 Dover, New York 1961);

[8b] L. Boltzmann, *Vorlesungen über Gastheorie.* 2 vols. Leipzig: Barth, 1896, 1898. This book has been translated into English by S.G. Brush, *Lectures on Gas Theory*, (Cambridge University Press, London 1964).

[9] L. Boltzmann, On Zermelo's Paper "On the Mechanical Explanation of Irreversible Processes". *Annalen der Physik* **60**, 392–398 (1897).

[10] O. Lanford, Physica A **106**, 70 (1981).

[11] H. Spohn, *Large Scale Dynamics of Interacting Particles*, Springer-Verlag, New York (1991). A. De Masi, E. Presutti, *Mathematical Methods for Hydrodynamic Limits*, Lecture Notes in Math 1501, Springer-Verlag, New York (1991). J.L. Lebowitz, E. Presutti, H. Spohn, J. Stat. Phys. **51**, 841 (1988); R. Esposito and R. Marra, *J. Stat. Phys.* **74**, 981 (1994); C. Landim, C. Kipnis, *Scaling Limits for Interacting Particle Systems*, Springer-Verlag, Heidelberg, 1998.

[12] J.L. Lebowitz and H. Spohn, *Communications on Pure and Applied Mathematics*, XXXVI,595, (1983); see in particular section 6(i).

[13] L. Boltzmann, Ann. der Physik **57**, 773 (1896). Reprinted in 4h).

[14] R. P. Feynman, *The Character of Physical Law*, MIT Press, Cambridge, Mass. (1967), ch. 5.

[15] R. Penrose, *The Emperor's New Mind*, Oxford U.P., New York (1990), ch. 7.

[16] J. W. Gibbs, *Elementary Principles in Statistical Mechanics*, Dover, New York (1960). Reprint of the original 1902 edition.

[17a] M. E. Fisher, *Arch. Ratl. Mech. Anal.* **17**, 377 (1964).
[17b] D. Ruelle, *Statistical Mechanics*, Benjamin (1969).
[17c] R. B. Griffiths, in *Phase Transitions and Critical Phenomena*, Vol. 1, C. Domb and H. S. Green, eds., Academic Press, 1972 (see also other articles there).
[17d] R. Baxter, *Exactly Solvable Models in Statistical Mechanics*, Academic (1982).
[17e] Ya. G. Sinai, *Theory of Phase Transitions: Rigorous Results*, Pergamon (1982).
[17f] H. O. Georgii, *Gibbs Measures and Phase Transitions*, (W. de Gruyter, 1988).
[17g] R. Fernández, J. Fröhlich, and D. A. Sokal, *Random Walks, Critical Phenomena, and Triviality in Quantum Field Theory*, Springer, Berlin (1992).
[17h] B. Simon, *The Statistical Mechanics of Lattice Gases*, Princeton Univ. Press (1993).

[18] C. N. Yang and T. D. Lee, Statistical Theory of Equations of State and Phase Transitions. I. Theory of Condensation, *Phys. Rev.* **87**, 404–409 (1952); T. D. Lee and C. N. Yang, Statistical Theory of Equatins of State and Phase Transitions. II. Lattice Gas and Ising Model, *Phys. Rev.* **87**, 410–419 (1952).

[19] c.f. T. M. Liggett, *Interacting Particle Systems*, Springer (1985); T. Vicsek, *Fractal Growth Phenomena*, World Scientific. P. Meakin, *Fractals, Scaling and Growth Far from Equilibrium*, Cambridge University Press.

[20] J. L. Lebowitz and A. Martin-Löf, *Commun. Math. Phys.* **25**, 276 (1972).

[21] C. M. Fortuin, P. W. Kasteleyn, and J. Ginibre, *Commun. Math. Phys.* **22**, 89 (1971).

[22] R. Peierls, On Ising's Model of Ferromagnetism, *Proc. Camb. Phil. Soc.* **32**, 477–481 (1936).

[23] R. L. Dobrushin and S. B. Shlosman, in *Statistical Physics and Dynamical Systems*, J. Fritz, A. Jaffe, and D. Szasz eds. (Birkenhauser, N.Y., 1985), pp. 347–370; 371–403; R. L. Dobrushin, J. Kolafa and S. B. Shlosman, *Comm. Math. Phys.* **102**, 89–103 (1985); D. C. Radulescu and D. F. Styer, *Jour. Stat. Phys.* **49**, 281 (1987).

[24] R. B. Griffiths, *J. Math. Phys.* **8**, 478 (1967); D. G. Kelley and S. Sherman, *J. Math. Phys.* **9**, 466 (1969).

[25] R. L. Dobrushin, *Th. Prob. Appl.* **17**, 382 (1972).

[26] M. Aizenman, *Phys. Rev. Let.* **43**, 407 (1979); *Comm. Math. Phys.* **73**, 83 (1980).

[27] H. van Beijeren, *Phys. Rev. Lett.* **38**, 993 (1977); H. van Beijeren, *Comm. Math. Phys.* **40**, 1 (1975).

[28] L. Onsager, *Phys. Rev.* **65**, 117 (1944); for some rigorous results on other two dimensional modes see J. Fröhlich and T. Spencer, Kosterlitz-Thouless Transition in the Two-Dimensional Plane Rotator and Coulomb Gas, *Phys. Rev. Lett.* **46**. 1006–1009 (1981).

[29] M. Aizenman, Proof of the Triviality of $\phi_d^4$ Field Theory and Some Mean-Field Features of Ising Models for $d > 4$, *Phys. Rev. Let.* **47**, 1–4 (1981).

[30] J. L. Lebowitz, *Jour. Stat. Phys.* **16**, 463 (1976).

[31] P. W. Anderson and G. Yuval, *Phys. Rev. Lett.* **23**, 89 (1969); *J. Phys. C* **4**, 607 (1971); F. J. Dyson, *Comm. Math. Phys.* **12**, 91 (1969); **21**, 269 (1971); M. Aizenman, J. T. Chayes, L. Chayes, and C. M. Newman, *J. Stat. Phys.* **50**, 1 (1988).

[32] S. A. Pirogov and Ya. G. Sinai, Phase Diagrams of Classical Lattice Systems, *Theor. and Math. Phys.* **25**, 358–369, 1185–1192 (1975); E. I. Dinaburg and Ya. G. Sinai, Contour Models with Interaction and Their Applications, *Sel. Math. Sov.* **7**, 291–315 (1988); R. L. Dobrushin and M. Zahradnik, Phase Diagrams for Continuous Spin Models. Extention of Pirogov-Sinai Theory, in *Mathematical Problems of Statistical Mechanics and Dynamics*, R. L. Dobrushin, ed., Dordrecht, Boston: Kluwer Academic Publishers, 1–123 (1986); R. Kotecky and D. Preiss, Cluster Expansion for Abstract Polymer Models, *Comm. Math. Phys.* **103**, 491–498 (1986).

[33] R. L. Dobrushin, *Func. Anal. Appl.* **2**, 31 (1968).

[34] F. J. Dyson and A. Lenard, *J. Math. Phys.* **8**, 423 (1967); A. Lenard and F. J. Dyson, *J. Math. Phys.* **9**, 689 (1968); J. L. Lebowitz and E. H. Lieb, *Phys. Rev. Lett.*

**22**, 631 (1969); E. H. Lieb and J. L. Lebowitz, *Adv. Math.* **9**, 316 (1972); E. H. Lieb, *Rev. Mod. Physics* **48**, 553 (1976);

[35] M. E. Fisher, Condensed Matter Physics: Does Quantum Mechanics Matter?, in *Niels Bohr: Physics and the World*, Eds. H. Feshbach, T. Matsui and A. Oleson, Harwood Academic Publ., Chur, 65–115 (1988).

[36] D. Ruelle, Existence of a Phase Transition in a Continuous Classical System, *Phys. Rev. Let.* **27**, 1040–1041 (1971).

[37] B. Widom and J. S. Rowlinson, New Model for the Study of Liquid-Vapor Phase Transitions, *J. Chem. Phys.* **52**, 1670–1684 (1970).

[38] B. U. Felderhof and M. E. Fisher, Phase Transitions in One-Dimensional Cluster-Interaction Fluids, IA. Thermodynamics, IB. Critical Behavior, II Simple Logarithmic Model, *Annals of Phys.* **58** N1, 176–216, 217–267, 268–280 (1970); K. Johansson, On Separation of Phases in One-Dimensional Gases, *Comm. Math. Phys.* **169**, 521–561 (1995).

[39] J. L. Lebowitz, A. Mazel and E. Presutti, Rigorous Proof of a Liquid-Vapor Phase Transition in a Continuum Particle System, *Phys. Rev. Lett.* **80**, 4701–4704 (1998); J. L. Lebowitz, A. Mazel and E. Presutti, Liquid-Vapor Phase Transitions for Systems with Finite Range Interactions, *Jour. Stat. Phys.*, to appear, Los Alamos archive, cond-mat/9809144, Texas archive 98-594.

[40] M. Kac, G. Uhlenbeck and P. C. Hemmer, On the Van der Waals Theory of Vapor-Liquid Equilibrium, *J. Mat. Phys.* **4**, 216–228, 229–247 (1963); *J. Mat. Phys.* **5**, 60–74 (1964).

[41] J. L. Lebowitz and O. Penrose, Rigorous Treatment of the Van de Waals Maxwell Theory of the Liquid-Vapor Transition, *J. Mat. Phys.* **7**. 98–113 (1966).

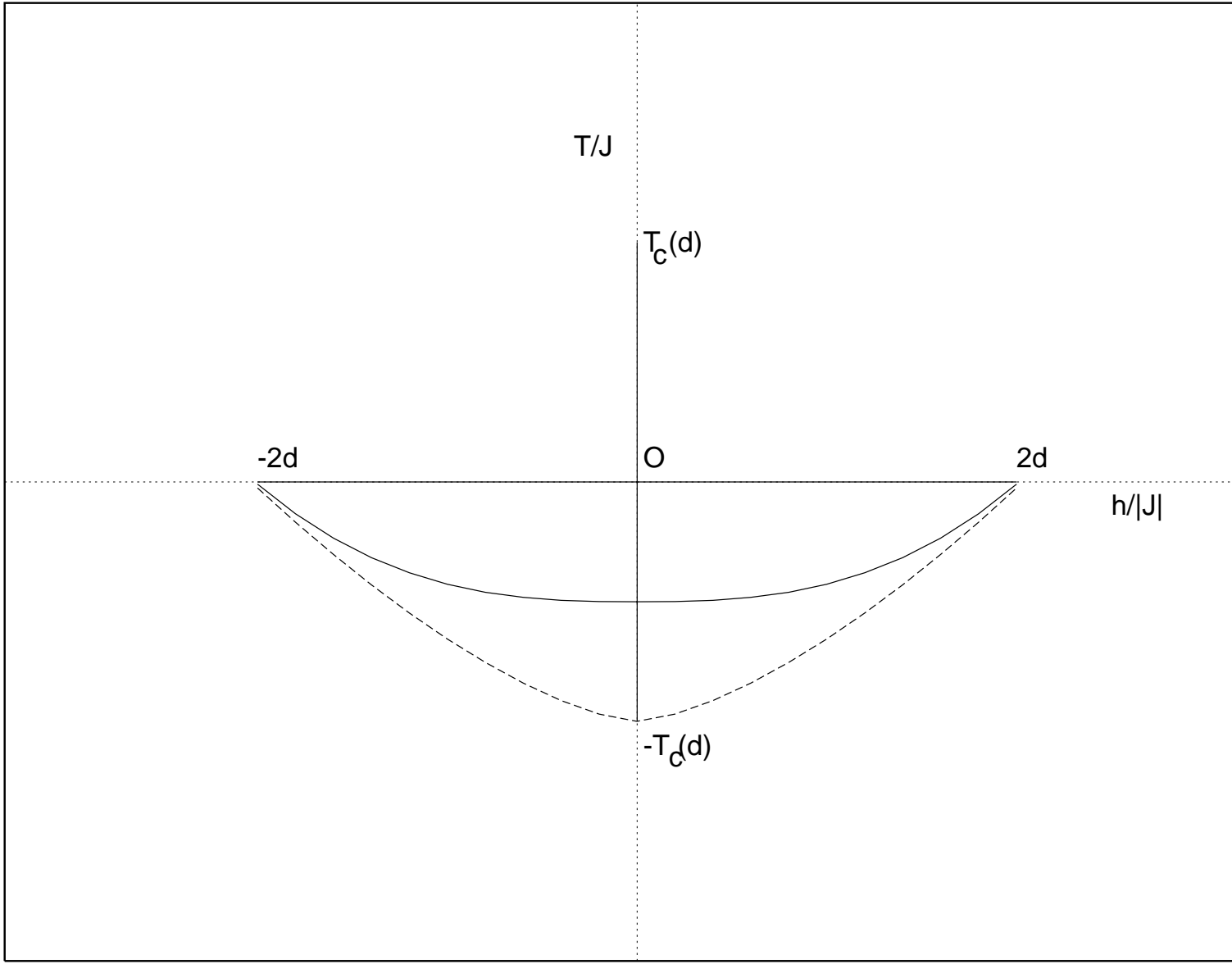

FIG. 1. Schematic phase diagram of nearest neighbor Ising model on a simple cubic lattice in dimensions $d \geq 2$. The ground states of the antiferromagnetic system are degenerate for $|h| \leq 2|J|d$. For $d = 1$, $T_c = 0$.